
%
%

\input  PHYZZX
\PHYSREV
\REF\Davis{R.~Davis, Gen. Rel. Grav. {\bf 19}, 331 (1987).}
\REF\Turok{N.~Turok, Phys. Rev. Lett. {\bf 63}, 2625 (1989).}
\REF\Vilenkin{M.~Barriola and
A.~Vilenkin, Phys. Rev. Lett. {\bf 63}, 341 (1989).}
\REF\Benneti{D.~Bennett and S.~Rhie, Phys. Rev. Lett. {\bf 65}, 1709
(1990).}
\REF\Paninal{L.~Perivolaropoulos, BROWN--HET--775 (November 1990).}
\REF\Bennetii{D.~Bennett and S.~Rhie, UCRL--IC--104061 (June 1990).}
\REF\Dine{M.~Dine and N.~Seiberg, Nucl. Phys. {\bf B301}, 357
                  (1988).}
\REF\Derend{J.~P.~Derendinger, L.~E.~Ib\'a\~nez, and
    H.~P.~Nilles, Phys. Lett. {\bf 155B}, 65 (1985); M.~Dine, R.~Rohm,
    N.~Seiberg, and E.~Witten, Phys. Lett. {\bf 156B}, 55 (1985).}
\REF\Rey{S.-J.~Rey, {\it Axionic String Instantons and Their
    Low-Energy Implications,} Invited Talk at Tuscaloosa Workshop on
    Particle Theory and Superstrings, ed. L.~Clavelli and B.~Harm, World
  Scientific Pub., (November, 1989); Phys. Rev. D {\bf 43}, 526 (1991).}
\REF\Ferrara{S.~Ferrara, D.~L\"ust, A.~Shapere, and S.~Theisen,
    Phys. Lett {\bf 225B}, 363 (1989).}
\REF\Cvetic{M.~Cveti\v c, A.~Font, L.~E.~Ib\'a\~nez, D.~L\"ust,
    and F.~Quevedo, Nucl. Phys. {\bf B361} (1991) 194.
    }
\REF\Kaplun{V.~Kaplunovsky, Nucl. Phys. {\bf B307}, 145 (1988).}
\REF\Dixon{L.~Dixon, V.~Kaplunovsky, and J.~Louis,
Nucl. Phys. {\bf B355}, 649 (1991); J.~Louis, SLAC--PUB--5527
(April 1991).}
\REF\Font{A.~Font, L.~E.~Ib\'a\~nez, D.~L\"ust, and F.~Quevedo,
    Phys. Lett. {\bf 245B}, 401 (1990);
          S.~Ferrara, N.~Magnoli, T.~R.~Taylor, and
    G.~Veneziano, Phys. Lett. {\bf 245B}, 409 (1990);
           P.~Binetruy and M.~K.~Gaillard, Phys. Lett. {\bf
    253B}, 119 (1991).}
\REF\Nilles{H.~P.~Nilles and M.~Olechowski, Phys. Lett. {\bf
    248B}, 268 (1990).}
\REF\CV{M.~Cveti\v c, S.~J.~Rey and F.~Quevedo,``Stringy Domain Walls
and Target Space Duality'', UPR--0445-T (April 1991) to appear in
Phys. Rev. Lett.}
\REF\Kim{J.~E.~Kim, Phys. Rev. Lett. {\bf 43}, 103 (1979);
    M.~Dine, W.~Fischler, and M.~Srednicki, Phys. Lett. {\bf 104B}, 199
    (1981); and Nucl. Phys. {\bf B189}, 575 (1981); M.~B.~Wise,
    H.~Georgi, and S.~L.~Glashow, Phys. Rev. Lett. {\bf 47}, 402 (1981);
    A good review is by J.~E.~Kim, Phys. Rep. {\bf 150}, 1 (1987).}
\REF\Sikivie{P.~Sikivie, Phys. Rev. Lett. {\bf 48}, 1156
    (1982);  G. Lazarides and Q. Shafi, Phys. Lett.{\bf 115B} (1982) 21;
    For a review, see A.~Vilenkin, Phys. Rep. {\bf 121}, 263
    (1985).}
\REF\modularform{B. Schoeneberg, \sl Elliptic Modular Functions, \rm
Springer, Berlin-Heidelberg (1970);
J. Lehner, \sl Discontinuous Groups and
Automorphic Functions, \rm
 ed. by the American Mathematical Society,
(1964).}
\REF\vafa{
P. Fendley, S. Mathur, C. Vafa and N.P. Warner, Phys. Lett.
\bf B243\rm (1990) 257.}
\REF\ours{M. Cveti\v c, S. Griffies, and S.-J. Rey,
{\sl Physical Implications of Stringy Domain Walls,} UPR--471.}
\REF\flst{S. Ferrara, D. L\"ust, A. Shapere and S. Theisen, Phys.
Lett. {\bf B225},     (1989) 363.}
\REF\bigpaper{M. Cveti\v c, A. Font, L.E. Ib\'a\~nez, D. L\"ust and
F. Quevedo, \sl Target Space Duality, Supersymmetry Breaking and the
Stability of Classical String Vacua, Nucl. Phys. \bf B361 \rm (1991)
194.}
\REF\domainansatz{A. Vilenkin, Phys. Lett. \bf 133B \rm (1983) 177:
J. Ipser and P. Sikivie, Phys. Rev. \bf D30 \rm (1984) 712.}
\REF\positiveenergy{E. Witten, Comm. Math. Phys. \bf 80 \rm (1981) 381.}
\REF\falsevacuum{S. Coleman, Phys. Rev. \bf D15 \rm (1977) 2929:
C. Callan and S. Coleman, Phys. Rev. \bf D16 \rm (1977) 1762.}
\REF\CQRprep{M.Cveti\v c, F.~Quevedo, and S.-J.~Rey, in preparation.}
\REF\CLO{M. Cveti\v c, J. Louis, and B. Ovrut,        Phys. Lett.
    {\bf 206B}, 227 (1988) and Phys. Rev. D {\bf 40}, 684 (1989).}
\REF\Gilm{R.~Gilmore, Lie Groups, Lie Algebras, and Some of Their
     Applications, John Wiley and Sons, (1974).}
\REF\Skyr{Skyrme, G. Adkins, C. Nappi, and E. Witten, Nucl. Phys.
     {\bf B228}, 5521 (1983).}
\nopubblock
\titlepage
\line{\hfill UPR--485--T
}
\line{\hfill September 1991
}
\title{Topological Defects in the Moduli Sector of String Theory\foot{
Talk Presented at Strings `91 Workshop, May 20--25, 1991, Stony Brook,
N.Y.}}
\author{Mirjam Cveti\v c}
\address{
Department of Physics\break
University of Pennsylvania\break
Philadelphia, PA 19104--6396\break}
\vfill
\singlespace
\vsize=630pt
\abstract{
We point out that the moduli sector of the $(2,2)$ string
compactification with its nonperturbatively preserved non-compact
         symmetries is a fertile framework to study global topological
defects, thus providing a natural source for the large scale structure
formation. Based on the target space modular invariance of the
nonperturbative superpotential of the four-dimensional $N=1$
supersymmetric string vacua, topologically stable stringy domain walls
are found.  They are supersymmetric solutions, thus saturating the
Bogomolnyi bound.
It is also shown that
there are moduli sectors that allow for the global monopole-type
and texture-type
configurations whose radial stability is ensured by higher derivative
terms.}
\medskip
Topological defects occur during the spontaneous break-down of gauge
symmetries, as a consequence of the nontrivial homotopy group $\Pi_n$ of
the vacuum manifolds.  Their existence has important cosmological
consequences.  In particular global topological defects, like
textures\refmark{\Davis,\Turok}\ and more recently, global
monopoles\refmark{\Vilenkin,\Benneti}\ as well as global $\Pi_2$
textures\refmark{\Paninal,\Bennetii}\
were proposed as a source of large scale
structure formation.  On the other hand, in the framework of
grand-unified theories (or theories beyond the standard model) it is
often unnatural to impose global non-Abelian symmetries that would
ensure the existence of such global topological defects.
Here we would like to point out that in the string theory, the moduli
sector of $(2,2)$ string compactification provides a natural framework
for such global defects, with its potentially important physical
implications.

In $(2,2)$ string compactifications, where $(2,2)$ stands for $N=2$
left-moving as well as $N=2$ right-moving world-sheet supersymmetry,
there are massless fields -- moduli $M$ -- which have no potential, {\it
i.e.} $V(M)\equiv0$, to all orders in string loops\rlap.\refmark{\Dine}\
Thus perturbatively there is a large degeneracy of string vacua, since
any vacuum expectation value of moduli corresponds to the vacuum
solution.  On the other hand it is known that nonperturbative stringy
effects like gaugino condensation\refmark{\Derend}\ and axionic string
instantons\refmark{\Rey}\ give rise to the nonperturbative superpotential.

In the case of the modulus $T$ associated with the internal size of the
compactified space for the so-called flat background compactifications
({\it e.g.}, orbifolds, self-dual lattice constructions, fermionic
constructions) the generalized target space duality is characterized by
noncompact discrete group $PSL(2,Z)=SL(2,Z)/Z_2$ specified by
$T\rightarrow{{aT-ib}\over{icT+d}}$ with $a$, $b$, $c$, $d\in Z$ and
$ad-bc=1$.  If one assumes that the generalized target space duality is
preserved even nonperturbatively\rlap,\refmark{\Ferrara,\Cvetic}\
the form of
the nonperturbative superpotential is very
restrictive\rlap.\refmark{\Cvetic}\
The fact that this is an exact symmetry
of string theory even at the level of nonperturbative effects is
supported by genus-one threshold
calculations\rlap,\refmark{\Kaplun,\Dixon}\
which in turn specify the form of the gaugino
condensate\rlap.\refmark{\Font,\Nilles}\

This phenomenon has intriguing physical implications leading to the
stable supersymmetric domain walls\rlap.\refmark{\CV}\  This physics of
modulus $T$ is actually a generalization of the well known axion
physics\refmark{\Kim}\ introduced to solve the strong $CP$ problem in
QCD.  Spontaneously broken global $U(1)$ Peccei-Quinn symmetry is
non-linearly realized through a pseudo-Goldstone boson, the invisible
axion $\theta$.  Nonperturbative QCD effects through the axial anomaly
break explicitly $U(1)$ symmetry down to $Z_{N_f}$, by generating an
effective potential proportional to $1-\cos N_f\theta$.  This potential
leads to domain wall solutions \refmark{\Sikivie}\
with $N_f$ walls meeting at the axionic
strings\rlap.\refmark{\Kim}\

As an instructive example let's
first consider a global supersymmetric theory by with
$$
L = G_{T \bar T} |\nabla T|^2 + G^{T \bar T} |\partial_T W(T)|^2
\eqno (1)
$$
Here,
$ G_{T \bar T} \equiv \partial_T \partial_{\bar T} K(T, \bar T)\  $
is the positive definite metric on the complex modulus space and
the superpotential, $W$,
is a rational polynomial $ P(j(T))$
of the modular-invariant
function $j(T)$\refmark{\modularform}\
{\it i.e.}
a modular invariant
form of $PSL(2,\bf Z)$.
The potential
$ V\equiv G^{T \bar T} |\partial_T W(T)|^2
= G^{T \bar T}
|\partial_j  P(j) \partial_T j(T)|^2$
has at least
two isolated zeros
at $T = 1$ and $T=\rho\equiv e^{i \pi/6}$
in the fundamental domain $\cal D$ for $T$, {\it i.e.} when $|\partial_T
j(T)|^2=0.$\refmark{\modularform}\
Other isolated degenerate minima might as well arise when
$|\partial_j   P(j)|^2=0$.

Then, the mass per unit area of the domain wall can be written
as:\refmark{\vafa}
$$
\mu = \int_{-\infty}^{\infty} dz\, G_{T \bar T}
| \partial_z  T - e^{i\theta} G^{T \bar T} \partial_{\bar T}
\bar W(\bar T)|^2  + 2 Re (e^{- i \theta} \Delta W)
\eqno (2)
$$
where
$\Delta W \equiv  W(T(z = \infty)) - W(T(z= -\infty))$. The arbitrary
phase
$\theta$ has to  be chosen such that
 $e^{i \theta} = \Delta W / |\Delta W|$,
thus maximizing the cross term in Eq. (2).
Then, we find $\mu \ge K \equiv 2 |\Delta W|$, where $K$ denotes the kink
number. Since $\partial_T W$
is analytic in $T$, the line integral over $T$ is \sl path independent \rm
as for a conservative force. The minimum is obtained only if
the Bogomolnyi bound $\partial_z T(z) = G^{T \bar T} e^{i \theta}
\partial_{\bar T} \bar W(\bar T(z))$
is saturated. In this case
$
\partial_z W(T(z)) = G^{T \bar T} e^{i \theta} |\partial_T W(T(z))|^2
$,
which implies that the phase of $\partial_z W$ does not change with $z$.
Thus, the supersymmetric domain wall is a mapping from the
 z-axis $[-\infty,
\infty]$ to a \sl straight line \rm  connecting between two degenerate
vacua in the $W$-plane.
We would like to emphasize that this result is general; it
applies to any globally supersymmetric theory with disconnected
degenerate minima that preserve supersymmetry.

For the superpotential, {\it e.g.}
$W (T) = (\alpha')^{-3/2} j(T)$
 the potential has two isolated  degenerate
minima at $T = 1$ and $T=\rho\equiv e^{i \pi/6}$.
 At these fixed points,
$j(T=\rho) = 0$ and $j(T=1) = 1728$. Therefore, the mass per unit area
is $\mu = 2 \times 1728 (\alpha')^{-3/2}$.
Other cases can be worked out analogously.\refmark{\ours}\

The case with  gravity restored
has a K\"ahler potential $K=-3 \log (T+\bar T)$ and the superpotential
should transform
as a weight $-3$
 modular function under modular
 transform\-a\-ti\-ons.\refmark{\flst,\Cvetic}
The most general choice, nonsingular everywhere in the fundamental
domain $\cal D$, is
$$
W_{m,n} (T) = { H_{m,n} (T) \over \eta (T)^6}, \,\,\,\,
H_{m,n} \equiv (j (T) - 1728)^{m/2} \cdot j^{n/3} (T) P  (j(T)),
\,\,\,m,n = {\bf R}^+
\eqno (3)
$$
Here, $\eta(T)$ is the Dedekind eta function, a modular form of weight
$1/2$ and $P (j(T))$ is an arbitrary
polynomial of $j(T)$. The potential is of the following form:
$$
V_{m,n} (T, \bar T) =  { 3 |H|^2 \over (T + \bar T)^3 |\eta|^{12}}
( |{(T + \bar T) \over 3} ({\partial_T H \over H} +
 {3 \over 2 \pi} \hat G_2) |^2 -1)
\eqno (4)
$$
where
$
\hat G_2=-{4\pi}\partial_T \eta/{\eta} -2\pi/(T+\bar T)$.
In general the scalar potential (4) has
an anti-de Sitter minimum with broken supersymmetry\rlap.
\refmark\Cvetic\
However, one can see that for
$m\geq 2,\  n\geq 2$ and $ P(j)=1$, the potential is semi-positive
definite with the two isolated minima at $T=1$ and $T=\rho$
with {it unbroken local   supersymmetry}  just like in the global
supersymmetric case.

 We now minimize the domain wall mass density. By the planar
 symmetry, the most general {\it static}
 Ansatz  for  the metric\refmark{\domainansatz}\   is
 $ ds^2 = A(|z|) (-dt^2 + d z^2) + B(|z|) (dx^2 + dy^2)$
 in which the domain wall is oriented parallel to $(x,y)$ plane.
Using the supersymmetry transformation laws
$$\eqalign{
\delta \psi_{\mu\alpha}
 &= [\nabla_\mu(\omega)  -{ i \over 2} Im
(G_T \nabla_\mu T)] \epsilon_{\alpha}  + {1 \over 2}
(\sigma_\mu \bar \epsilon)_\alpha  e^{G \over 2},\cr
\delta \chi_\alpha  &= {1 \over 2}(\sigma^\mu \bar \epsilon)_\alpha
\nabla_\mu T
 - e^{G \over 2}
G^{T \bar T} G_{\bar T} \epsilon_\alpha}\eqno (5)
$$
with commuting, covariantly constant, chiral spinors $\epsilon_\pm$,
the ADM mass density $\mu$ can be expressed as\refmark{\positiveenergy}
$$
\mu \mp K =  \int \! dz \, \sqrt g [ g_{ij}
\delta \psi^{\dagger i}\delta \psi^j + {1\over 2} G_{T\bar T}
\delta \chi^\dagger \delta \chi] \ge 0.
\eqno (6)
$$
The $i,j$ indices are for spatial directions.
The minimum of the Bogomolnyi bound is achieved if Eq.(5) vanish.
Again, the stringy domain wall is stabilized by the topological kink number.

Unfortunately, the nice holomorphic structure of the scalar potential is
lost. In other words, there is now
 a \sl holomorphic anomaly \rm in the
scalar potential due to the supergravity coupling.
This implies that the path connecting two
degenerate vacua in superpotential space is \sl not \rm  a straight line.
In fact, one can understand the motion as a {\it geodesic}
 path in a nontrivial
K\"ahler metric, thus in $G(T, \bar T)$.
One can show (numerically) that
in  our example the path along the circle $T= \exp{i\theta(z)}$ with
$\theta= (0, \pi/6)$, $i. e.$ , the
 self-dual line of $T \rightarrow 1/T$ modular transformation, is the \sl
geodesic \rm
path connecting between $T=1$ and $T=\rho$ in the scalar potential space.
Thus, we have again established an existence of stable domain walls.
The superpotential
is quite complicated, however,
the numerical solution in can be obtained.\refmark{\ours}

It is interesting to note that  stringy cosmic strings\foot{It is
intriguing that the present kink solitons also appear in integrable,
supersymmetric two-dimensional $N=2$ Landau-Ginzburg
models\refmark{\vafa}}\
can be viewed as boundaries of our domain walls. Because the domain wall
number is two, the intersection of two such domain walls is precisely
the  line of stringy cosmic strings. On the other hand such stable
domain walls are disastrous from the cosmological point of view.
One possible solution to this problem  is that after supersymmetry
breaking, the degeneracy of the two minima is lifted. In that case,
the domain wall becomes unstable via the false vacuum
 decay.\refmark{\falsevacuum}

We would now like to point out\refmark{\CQRprep}
 the existence of other global
topological defects, like global monopole-type
 and texture-type defects in the moduli sector of string theory.
 Such defects could exist in the study of the symmmetry structure
of the effective theory when there are more than one modulus (which of
course is a generic situation).  We shall illustrate the idea using
examples based on the so called flat backgrounds, {\it i.e.}
generalization of $SL(2,{\bf Z})$.

For that purpose we shall study the simplest example of $Z_4$ manifold
with continuous symmetry $SU(2,2)$ on the four moduli
$$
\bf T\equiv\left[\matrix{
T_{11} & T_{12} \cr
T_{21} & T_{22} \cr}\right]\eqno(7)$$
of compactified space.  Note that the moduli $\bf T$ live on the coset
$SU(2,2)/SU(2)\times SU(2)\times U(1)$.  The continuous non-compact
symmetry $SU(2,2)$ is an {\it exact}
symmetry\refmark{\CLO} at least at the
string tree-level.  Note that this continuous symmetry in the modulus
could be broken down to the discrete subgroup $SU(2,2,Z)$ due to
nonperturbative effects, {\it e.g.} gaugino condensation and/or axionic
instanton effects.  However, at this point we shall stick to the
continuous symmetry.
For the time being we shall keep in mind that $SU(2,2,Z)$ is
the vacuum symmetry and thus the $\bf T$ fields should live in the
fundamental domain of $SU(2,2,Z)$.

The maximal comapct symmetry of $SU(2,2)$ is $SU(2)_A\times
SU(2)_B\times U(1)\subset SU(2)_{A+B}$.  Note also that in projective
coordinates\rlap:\refmark{\Gilm}\
$  \bf Z=(1-\bf T)/(1+\bf T)
.$
$\bf Z$ transforms as
${\bf 1}+{\bf 3}$ under $SU(2)_{A+B}$.  The ansatz
${\bf Z}=\sum_{a=1}^3\sigma_a V_a$
with $V_a=f(r)x_a/r$ ensures the map of
$\bf Z$ on the $S^2$.

 Let us concentrate now on the
 Lagrangian for the $\bf Z$ field and thus a
specific solution for $f(r)$.
Note that the $\bf Z$ fields have {\it no}
potential to all orders in string loops.  Thus the kinetic energy
term\refmark{\CLO}\
shrinks $f\rightarrow0$ due to Derrick's theorem                and thus
should be stabilized by higher derivative terms.
Such higher derivative terms
arise  even at the tree level of the
string theory.  They should respect the
noncompact $SU(2,2)$ symmetry.  Also, if one sticks to terms with at most
two time derivatives, one has a unique form for the terms that involve
four derivatives,
which is very similar
in nature to the Skyrme term\refmark{\Skyr} in the
Skyrmion model and can serve the same role as the stabilizing term.
In this case
$f=Cr$ as $r\rightarrow0$ and       $f=D/r^2$ as
$r\rightarrow\infty$
The energy stored in such a configuration is finite.
This is different from the standard global monopole
configuration\rlap,\refmark{\Vilenkin}\
 which has $f\rightarrow f_0$ as
$r\rightarrow\infty$ which has linearly divergent energy and thus long
range interaction relevant for large scale formation.

Another interesting observation would
be to study the texture-type configurations,
 which have a chance to occur within
this sector.  Namely, the $\bf Z$
fields transform as $\bf 4$ under the
compact symmetry $SU(2)_A\times SU(2)_B\sim SO(4)$ and thus the ansatz:
${\bf Z}=a(r)+b(r)\sum_{a=1}^3\sigma_a x_a/r$
is mapped onto $S^3$.  The potential problem in this case is an
impossibility to ensure $a^2(r)+b^2(r)=f^2$ where $f$ is a constant.
Interestingly,
$\left\{a(r),b(r)\right\}\rightarrow0$ as $r\rightarrow\infty$ and thus
the knot configuration disappears at large distances.

The above
studied  configurations are much milder defects than strings
and domain walls and they  have
finite range and thus finite energy.  Further study
of cosmological implications of such global defects is under
consideration.

I would like to thank my collaborators S.~Griffies, F.~Quevedo, and
S.-J.~Rey, for many fruitful discussions and enjoyable collaborations.
I would also like to thank the Aspen Center for Physics, the
International Centre for Theoretical Physics, Trieste, and CERN for
their hospitality.  The work is supported in part by the U.S. DOE Grant
DE--22418--281 and by a grant from
University of Pennsylvania Research Foundation and by the NATO
Research Grant \#900--700.
\refout\end